\documentclass{aa}
\usepackage{txfonts}
\usepackage{graphicx}

\newcommand\teff{$ {\rm T_{eff}}$}

\newcommand\loghe{${\rm \log{\frac{n_{He}}{n_{H}}}}$}

\newcommand{\Msolar}{\mbox{\,$\rm M_{\odot}$}}        
\begin{document}

\title{The Hottest Horizontal-Branch Stars in $\omega$ Centauri --
  Late Hot Flasher vs. Helium Enrichment\thanks{Based on observations
  with the ESO Very Large Telescope at Paranal Observatory, Chile
  (proposal IDs 075.D-0280(A) and 077.D-0021(A))}}
\author{S. Moehler\inst{1}
\and S. Dreizler\inst{2}
\and T. Lanz\inst{3}
\and G. Bono\inst{4}
\and A.V. Sweigart\inst{5}
\and A. Calamida\inst{4}
\and M. Monelli\inst{6}
\and M. Nonino\inst{7}}
\institute{
European Southern Observatory, Karl-Schwarzschild-Str. 2, D 85748
  Garching, Germany {\sl e-mail: smoehler@eso.org}
\and Georg-August-Universit\"at, Institut f\"ur Astrophysik,
Friedrich-Hund-Platz 1, D 37077 G\"ottingen, Germany {\sl e-mail: 
  dreizler@astro.physik.uni-goettingen.de} 
\and Department of Astronomy, University of Maryland, College Park, MD  
20742-2421, USA {\sl e-mail: lanz@astro.umd.edu}
\and INAF - Rome Astronomical Observatory, via Frascati 33, 00040 
Monte Porzio Catone, Italy {\sl e-mail: bono,calamida@mporzio.astro.it}
\and NASA Goddard Space Flight Center, Code 667, Greenbelt, MD 20771,
  USA {\sl e-mail: Allen.V.Sweigart@nasa.gov}
\and IAC - Instituto de Astrofisica de Canarias, Calle Via Lactea,
E38200 La Laguna, Tenerife, Spain {\sl e-mail: monelli@iac.es}  
\and INAF - Trieste Astronomical Observatory, via G.B. Tiepolo 11,
40131 Trieste, Italy {\sl e-mail: nonino@ts.astro.it}
}

\abstract
{UV observations of some massive globular clusters uncovered a
significant population of very hot stars below the hot end of the
horizontal branch ({\bf HB}), the so-called blue hook stars. This
feature might be explained either as results of the late hot flasher
scenario where stars experience the helium flash while on the white
dwarf cooling curve or by the progeny of the helium-enriched
sub-population recently postulated to exist in some clusters. Previous
spectroscopic analyses of blue hook stars in $\omega$ Cen and NGC~2808
support the late hot flasher scenario, but the stars contain much less
helium than expected and the predicted C, N enrichment could not be
verified.}
{We want to compare effective temperatures, surface gravities, and
 abundances of He, C, and N of blue hook and canonical extreme
 horizontal branch ({\bf EHB}) star candidates to the predictions of
 the two scenarios.}
{Moderately high resolution spectra of stars at the hot end of the
  blue HB in the globular cluster $\omega$ Cen were analysed for
  atmospheric parameters and abundances using LTE and Non-LTE model
  atmospheres.} 
{In the temperature range 30,000\,K to 50,000\,K we find that 35\% of
  our stars are helium-poor (\loghe $< -2$), 51\% have solar helium
  abundance within a factor of 3 ($-1.5 \lesssim$ \loghe $\lesssim -0.5$) and
  14\% are helium-rich (\loghe $> -0.4$). We also find carbon
  enrichment in step with helium enrichment, with a maximum carbon
  enrichment of 3\% by mass.  }
{At least 14\% of the hottest HB stars in $\omega$~Cen show helium
  abundances well above the highest predictions from the helium
  enrichment scenario ($Y$ = 0.42 corresponding to \loghe $\approx
  -0.74$). In addition, the most helium-rich stars show strong carbon
  enrichment as predicted by the late hot flasher scenario. We
  conclude that the helium-rich HB stars in $\omega$ Cen cannot be
  explained solely by the helium-enrichment scenario invoked to
  explain the blue main sequence.}  \keywords{Stars: horizontal branch
  -- Stars: evolution -- Techniques: spectroscopic -- globular
  clusters: individual: NGC~5139} \maketitle

\section{Introduction}
\label{sec:intro}
UV-Visual colour-magnitude diagrams of the two very massive globular
clusters, $\omega$ Cen and NGC~2808, show a rather puzzling
``hook-like'' feature at the hot end of their extended horizontal
branches with stars lying below the canonical horizontal branch
(Whitney et al. \cite{whro98}; D'Cruz et al. \cite{dcoc00}; Brown et
al. \cite{brsw01}). These stars cannot be explained within the
framework of canonical stellar evolution.  Brown et al.\
(\cite{brsw01}) have proposed a ``flash-mixing'' scenario to explain
the blue hook stars.  According to this scenario stars which lose an
unusually large amount of mass will leave the red giant branch ({\bf
RGB}) before the helium flash and will move quickly to the
(helium-core) white dwarf cooling curve before igniting helium
(Castellani \& Castellani \cite{caca93}; D'Cruz et al.\ \cite{dcdo96};
Brown et al. \cite{brsw01}).  However, the evolution of these ``late hot
helium flashers'' differs dramatically from the evolution of stars
which undergo the helium flash on the RGB. Ordinarily when a star
flashes at the tip of the RGB or shortly thereafter, the large entropy
barrier of its strong hydrogen-burning shell prevents the products of
helium burning from being mixed to the surface.  Such canonical stars
will evolve to the zero-age horizontal branch ({\bf ZAHB}) without any
change in their hydrogen-rich envelope composition. In contrast, stars
that ignite helium on the white dwarf cooling curve, where the
hydrogen-burning shell is much weaker, will undergo extensive mixing
between the helium- and carbon-rich core and the hydrogen envelope
(Sweigart \cite{swei97}; Brown et al. \cite{brsw01}; Cassisi et
al. \cite{cas03}). Depending on where the helium flash occurs along
the white dwarf cooling curve, the envelope hydrogen will be mixed
either deeply into the core (``deep mixing'') or only with a
convective shell in the outer part of the core (``shallow
mixing''). In the case of deep mixing virtually all of the envelope
hydrogen is burned while in shallow mixing some of the envelope
hydrogen remains after the mixing phase (Lanz et al. \cite{labr04}).
One of the most robust predictions of the flash-mixing scenario is an
increase in the surface abundance of carbon to 3\% - 5\% (deep mixing)
or 1\% (shallow mixing) by mass. This increase is set by the carbon
production during the helium flash and is nearly independent of the
stellar parameters.  Nitrogen may also be enhanced due to the burning
of hydrogen on triple-$\alpha$ carbon during the flash-mixing
phase. For both deep and shallow mixing, the blue hook stars should be
helium-rich compared to the canonical EHB stars.

Alternatively, the recently observed split among the main sequence
stars of $\omega$ Cen and NGC\,2808 (Piotto et
al. \cite{pivi05,pibe07}) has been attributed to a sub-population of
stars with helium abundances as large as $Y$$\approx$0.4 (Norris
\cite{norr04}; D'Antona et al. \cite{dabe05}; D'Antona \& Ventura
\cite{dave07}; see Newsham \& Terndrup \cite{nete07} for cautionary
remarks). Lee et al. (\cite{lejo05}) have suggested that the blue hook
stars are the progeny of these proposed helium-rich main sequence
stars. If the blue hook stars were to be explained by the
helium-enrichment scenario, their helium abundance should not exceed
$Y$$\approx$0.4 and carbon should not be enriched at all.  Spectroscopic
observations of the blue (and supposedly helium-rich) main sequence
stars in $\omega$~Cen yield a carbon abundance of [C/M] = 0.0 (Piotto
et al. \cite{pivi05}).  This carbon abundance will decrease further as
the stars ascend the red giant branch, due to the extra-mixing process
that occurs in metal-poor red giants (Gratton et al. \cite{grsn00};
Kraft \cite{kraf94}).  Origlia et al. (\cite{orfe03}) have confirmed
that the RGB stars in $\omega$ Cen have the low $^{12}$C/$^{13}$C
ratios ($\approx$4) and low average carbon abundances ([C/Fe] =
$-$0.2) expected from this extra-mixing.  Thus the helium-enrichment
scenario predicts a carbon abundance by mass in the blue hook stars of
less than 0.1\%, i.e., at least a factor of 10 smaller than the carbon
abundance predicted by the flash-mixing scenario.

Previous spectra of the blue hook stars in $\omega$ Cen
(Moehler et al. \cite{mosw02}) and NGC~2808 (Moehler et al.\
\cite{mosw04}) showed that these stars are indeed both hotter and more
helium-rich than the canonical EHB stars. However, the blue hook stars
still show considerable amounts of hydrogen. Unfortunately due to
limited resolution and signal-to-noise ({\bf S/N}) we could not derive
good abundances for C and N. Instead we could only state that the most
helium-rich stars appear to show some evidence for C/N
enrichment. Therefore we started a project to obtain higher resolution
spectra of EHB and blue hook stars in $\omega$~Cen.

\section{Observations, Data Reduction, and Analysis}\label{sec:obs}

\begin{figure}[h!]
\vspace*{6.9cm}
\includegraphics{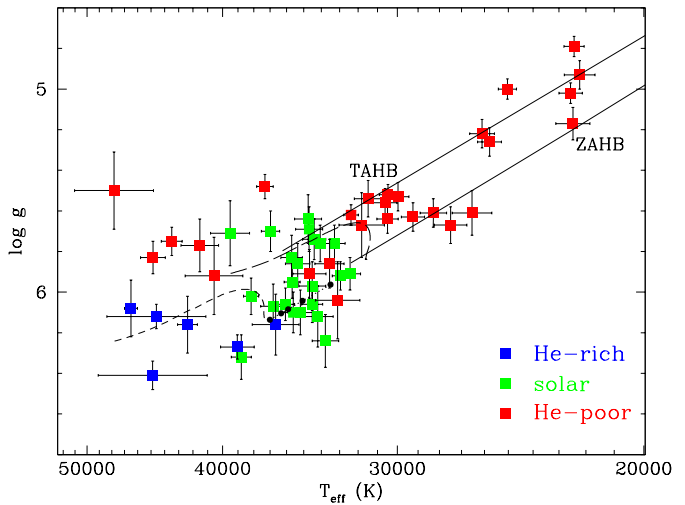}
\caption[]{Here we show the effective temperatures and surface
  gravities derived for our hottest target stars (formal errors only). 
  Helium-poor, solar helium, and helium-rich stars are marked by red,
  green, and blue squares, respectively. The
  solid lines mark the canonical HB locus for [M/H] = $-$1.5 from
  Moehler et al.  (\cite{mosw03}). The tracks for an early hot flasher
  (long-dashed line) and a late hot flasher (short-dashed line) show
  the evolution of such stars from the zero-age HB (ZAHB) towards
  helium exhaustion in the core (terminal-age HB = TAHB). The dotted
  line connects the series of ZAHB models computed by adding a
  hydrogen-rich layer to the surface of the ZAHB model of the late hot
  flasher.  The small dots mark -- with decreasing temperature --
  hydrogen layer masses of $0, 10^{-7}, 10^{-6}, 10^{-5},
  10^{-4}$\Msolar\ (for details see Moehler et
  al. \cite{mosw02}).}\label{Fig:Teff_logg}
\end{figure}

We selected stars along the blue HB in $\omega$ Cen from the
multi-band ($U,B,V,I$) photometry of Castellani et
al. (\cite{cast07}). These data were collected with the mosaic CCD
camera Wide Field Imager available at the 2.2m ESO/MPI telescope. The
field of view covered by the entire mosaic is $42'\times 48'$ across
the center of the cluster. These data together with multiband data
from the Advanced Camera for Surveys on board the Hubble Space
Telescope provided the largest sample of HB stars ($\approx$3,200)
ever collected in a globular cluster. Among them we concentrated on
the stars at the faint end of the HB, which are the most likely ``blue
hook'' candidates as shown by Moehler et al. (\cite{mosw02},
\cite{mosw04}). In order to avoid crowding problems, we only selected
isolated EHB stars. The astrometry was performed using the UCAC2
catalog (Zacharias et al. \cite{zacharias2004}), which does not cover
the central crowded regions. However, thanks to the large field
covered by current dataset the astrometric solution is based on
$\approx$3,000~objects with an rms error of 0\farcs06.

The spectroscopic data were obtained in 2005 (4 observations) and in
2006 (5 observations) in Service Mode using the MEDUSA mode of the
multi-object fibre spectrograph FLAMES$+$GIRAFFE on
the UT2 Telescope of the VLT. We used the low spectroscopic resolution
mode with the spectral range 3964\AA\ -- 4567\AA\ (LR2, R = 6400) and
observed spectra for a total of 101 blue hook and canonical blue
HB/EHB star candidates and 17 empty positions for sky
background. 

For our analysis we used the pipeline reduced data. For each exposure
we subtracted the median of the spectra from the sky fibres from the
extracted spectra. We corrected all spectra for barycentric motions.
The individual spectra of each target star have been cross-correlated
with appropriate template spectra, in order to search for radial
velocity variations. Since the few spectra per object did not permit a
sophisticated period search, we determined the standard deviation of
the radial velocity measurements for each star and compared it with
the S/N ratio of the spectra. As expected, the standard deviation
of the radial velocity measurements decreases with increasing S/N ratio.
None of our target stars deviates significantly from this
correlation, which would be the case for close binaries. 
Therefore none of our target stars appears to be in a close
binary system. After verifying that there were no radial velocity
variations we co-added all spectra for each star. The co-added and
velocity-corrected spectra were fitted with various model
atmospheres: metal-free helium-rich non-LTE (Werner \& Dreizler
\cite{wedr99}), metal-free helium-poor non-LTE (Napiwotzki
\cite{napi97}), and metal-rich helium-poor LTE (Moehler et
al. \cite{mosw00}) as described in Moehler et
al. (\cite{mosw04}). This procedure yielded the effective temperatures,
surface gravities, and helium abundances shown in Figs. 1 and 2. In
this paper we concentrate only on the hottest HB stars with \teff$>$20,000~K.

\section{Results and Discussion}\label{sec:Results}
\begin{figure}[h!]
\vspace*{6.9cm}
\includegraphics{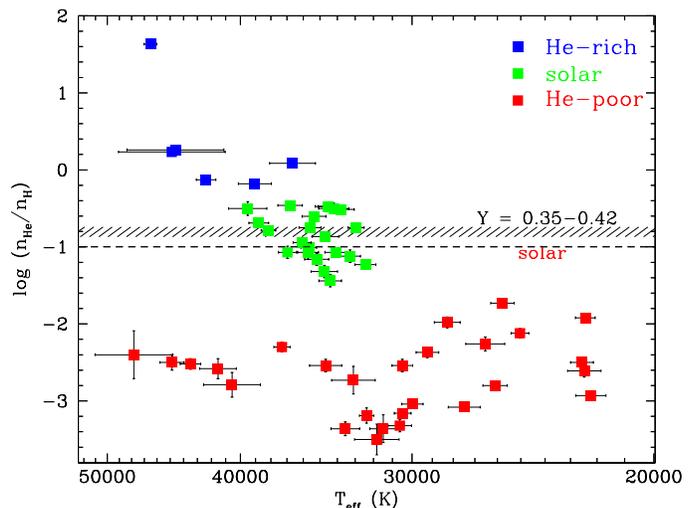}
\caption[]{Here we show the effective temperatures and helium
abundances for our hottest target stars (formal errors only). 
The dashed line marks solar helium abundance, the hashed area marks
the range for the helium-enrichment scenario.
The symbols have the same meaning as in 
Fig.~\ref{Fig:Teff_logg}.}\label{Fig:Teff_loghe}
\end{figure}

The helium-poor stars in Fig.~\ref{Fig:Teff_logg}
basically agree with the predictions of canonical evolutionary
theory in that they populate the HB up to its hot end and then also
contribute some evolved stars at higher effective temperatures and
lower surface gravities. As we move to hotter stars (\teff
$\gtrsim$30,000\,K), we find a clump of stars populating the range in
effective temperature and surface gravity between a fully mixed late
hot flasher and the hot edge of the canonical HB. These stars show
roughly solar helium abundance (cf. Fig.~\ref{Fig:Teff_loghe}). The
hottest stars lying along the evolutionary track of a fully mixed late
hot flasher show the highest helium abundances, albeit with still some
hydrogen in their atmospheres.  In the temperature range
30,000\,K to 50,000\,K we find that 35\% (15) of our stars are helium-poor
(\loghe $< -2$), 51\% (22) have solar helium abundance within a factor of 3
($-1.5 \lesssim$ \loghe $\lesssim -0.5$) and 14\% (6) are helium-rich
(\loghe $> -0.4$).

\begin{figure}[h!]
\vspace*{7.6cm}
\includegraphics{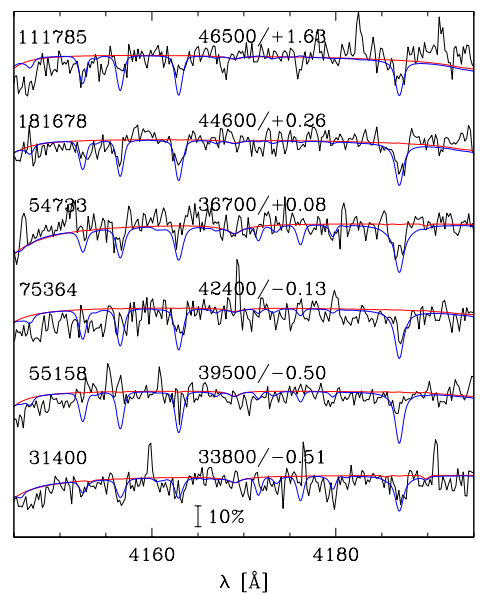}
\caption[]{Here we show sample spectra of stars with super-solar
  helium abundance, compared to model spectra with the
  cluster carbon abundance for metal-poor stars (red) and a
  carbon abundance of 3\% by mass (blue).  
  The labels give the number of the star, its effective
  temperature, and its helium abundance \loghe.}\label{Fig:carbon}
\end{figure}

The helium-rich stars also show evidence for carbon enrichment as
shown in Fig.~\ref{Fig:carbon}, unlike the hot (\teff $>$ 30,000\,K)
helium-poor stars where no \ion{C}{ii} and \ion{C}{iii} lines were
detected despite the higher S/N in their spectra.  We have constructed
additional TLUSTY NLTE line-blanketed model atmospheres (Hubeny \&
Lanz \cite{hula95}; Lanz~\& Hubeny \cite{lahu03,lahu07}) for the
atmospheric parameters of the helium-rich stars, assuming either
scaled-solar abundances appropriate for the dominant $\omega$~Cen
metallicity ([M/H]$=-$1.5) or the carbon- and nitrogen-rich abundances
predicted by the flash mixing scenario (mass fractions of 3\% and 1\%,
respectively). The comparison between observed and predicted
\ion{C}{ii} and \ion{C}{iii} lines indicates that the helium-rich
stars have a photospheric C mass fraction of at least 1\% and up
to 2--3\% for the stars with the strongest lines. The typical line
detection limit provides an upper limit of about 1\% by mass for the
N abundance. These C abundances represent a significant
enhancement relative to the expected C abundance in $\omega$~Cen
stars ($\lesssim$0.1\% at most for the most metal-rich stars).

Any discussion of the surface abundances in hot HB stars must consider
the effects of diffusion.  Fortunately the diffusion of H, He and the
CNO elements in the envelopes of stars following deep flash mixing has
been investigated by Unglaub (\cite{ungl05}).  Not surprisingly, the
results depend on the assumed mass loss rate and on the residual
hydrogen abundance remaining after the flash mixing.  For the low
residual hydrogen abundance $X$ = 0.0004 predicted by the Cassisi et
al. (\cite{cas03}) models, Unglaub (\cite{ungl05}) found that a star
will remain helium-rich with \loghe\ $\approx$0\ldots2 during most of
the HB phase, in rough agreement with the helium-rich stars in
Fig.~\ref{Fig:Teff_loghe}.  However, the residual hydrogen abundance
following flash mixing is quite uncertain, since it depends on the
mixing efficiency (Cassisi et al. \cite{cas03}) and possibly on where
the helium flash occurs along the white dwarf cooling curve.  For a
larger, but still low, residual hydrogen abundance of $X$ = 0.004,
Unglaub (\cite{ungl05}) found that diffusion can produce either a star
with near solar helium abundance or a star that is helium-poor by the
end of the HB phase.  The stars with roughly solar helium abundances
in Fig.~\ref{Fig:Teff_loghe} might be a consequence of such diffusion.
The ZAHB models with a hydrogen-rich layer in Fig.~\ref{Fig:Teff_logg}
show that the effective temperature will decrease as the amount of
surface hydrogen increases in qualitative agreement with the trend
towards lower effective temperatures between the helium-rich and solar
helium abundance stars in Fig. 2.  Unglaub (\cite{ungl05}) also noted
that the diffusion efficiency increases substantially once the helium
abundance approaches the solar value, leading to a rapid decrease in
\loghe\ and perhaps accounting for the gap between the solar helium
and helium-poor stars in Fig.~\ref{Fig:Teff_loghe}.  Diffusion in
flash-mixed stars also leads to a decrease in the carbon and nitrogen
abundances, which becomes more pronounced when the atmosphere is
hydrogen-rich. Thus the carbon abundances derived here for the
helium-rich stars may underestimate the initial carbon abundances in
these stars.  

The referee asked us to discuss the possibility that diffusion
may not be active in all stars above 30,000~K. Let us consider the
extreme case that only the helium-poor stars in this temperature range
are affected by diffusion. In this case the most helium-rich stars
could still be reconciled with the late hot flasher scenario, but the
same does not apply to the solar-helium stars. For the latter to be
considered as the progeny of the helium-enriched main sequence stars,
however, one would expect to see in Fig.~\ref{Fig:Teff_loghe} stars
highly concentrated at \loghe\ = $-0.82\ldots-0.74$
(i.e. $Y$=0.38\ldots0.42, Lee et al. \cite{lejo05}). Instead we see a
rather large scatter, esp. towards lower helium abundances. In
addition, the progeny of the helium-enriched main sequence should not
show strong carbon enhancements.

A puzzling effect becomes evident if one plots the spatial
distribution of our target stars: Dividing the sample along a
line running at 55$^\circ$ counter-clockwise from east-west, the helium-poor
stars are evenly distributed (28:30 for all, 8:7 for those above
30,000~K), while the stars with roughly solar or super-solar helium
abundance show a noticeable preference for the north-west section of
the globular cluster (17:5 and 5:1, respectively). This peculiar
spatial distribution appears similar to the reddening distribution
observed by Calamida et al. (\cite{cast05}), who found a clumpy
extinction variation with less reddened HB stars concentrated on the
east side of the cluster (see their Fig. 5).

\section{Conclusions}\label{sec:Conclusions}
All of these results taken together offer strong support for the late
hot flasher scenario as the explanation for the blue hook stars while
posing a significant problem for the helium-enrichment scenario.  This
scenario predicts helium enrichment of up to $Y = 0.35 \ldots 0.42$,
i.e. \loghe\ = $-$0.87 \ldots $-$0.74. However, 40\% -- 30\% of the
stars above 30,000\,K show helium abundances in excess of these
values, respectively. 
{\em This result together with the
observed carbon enhancement does not rule out the helium enhancement
scenario, but it implies that additional processes are required to
produce the hottest HB stars in $\omega$ Cen.}

\begin{acknowledgements}We thank the staff at the Paranal observatory and at
  ESO Garching for their excellent work, which made this paper
  possible. We also acknowledge that without the request from the ESO
  OPC to look into the data we have before applying again, these
  results would not have been found so soon. We thank the anonymous referee
 for his/her suggestions.
\end{acknowledgements}


\begin{thebibliography}{}
\bibitem[2001]{brsw01} 
Brown, T.~M., Sweigart, A.~V., Lanz, T., Landsman, W.~B., \& Hubeny,
I. 2001, ApJ, 562, 368 
\bibitem[2003]{cas03} 
Cassisi, S., Schlattl, H., Salaris, M., \& Weiss, A.  2003, ApJ, 582,
L43
\bibitem[2005]{cast05}
Calamida, A., Stetson, P. B., Bono, G., et al. 2005, ApJ, 634, L69 
\bibitem[1993]{caca93} 
Castellani, M., \& Castellani, V. 1993, ApJ, 407, 649\
\bibitem[2007]{cast07} 
Castellani, V., Calamida, A., Bono, G., et al. 2007, ApJ, 663, 1021
\bibitem[2005]{dabe05} 
D'Antona, F., Bellazzini, M., Fusi Pecci, F., Galleti, S., Caloi, V.,
\& Rood, R.~T. 2005, ApJ, 631, 868
\bibitem[2007]{dave07} 
D'Antona, F., Ventura, P. 2007, MNRAS, 379, 1431
\bibitem[1996]{dcdo96} 
D'Cruz, N.~L., Dorman, B., \& Rood, R.~T. 1996, ApJ, 466, 359
\bibitem[2000]{dcoc00} 
D'Cruz, N.~L., O'Connell, R.~W., Rood, R.~T., et al. 2000, ApJ, 530, 352
\bibitem[2000]{grsn00}
Gratton, R.~G., Sneden, C., Carretta, E., \& Bragaglia, A. 2000, A\&A, 354, 169
\bibitem[1995]{hula95}
Hubeny, I., \& Lanz, T. 1995, ApJ, 439, 875
\bibitem[1994]{kraf94}
Kraft, R. P. 1994, PASP, 106, 553
\bibitem[2004]{labr04} 
Lanz, T., Brown, T.~M., Sweigart, A.~V., Hubeny, I., \& Landsman,
W.~B. 2004, ApJ, 602, 342 
\bibitem[2003]{lahu03}
Lanz, T., \& Hubeny, I. 2003, ApJS, 146, 417
\bibitem[2007]{lahu07}
Lanz, T., \& Hubeny, I. 2007, ApJS, 169, 83
\bibitem[2005]{lejo05}
Lee, Y.-W., Joo, S.-J., Han, S.-I., et al. 2005, ApJ, 621, L57
\bibitem[2000]{mosw00} Moehler, S., Sweigart, A.~V., Landsman, W.~B.,
\& Heber, U. 2000, A\&A, 360, 120
\bibitem[2002]{mosw02} Moehler, S., Sweigart, A.~V., Landsman, W.~B.,
\& Dreizler, S. 2002, A\&A 395, 37
\bibitem[2003]{mosw03}
Moehler S., Landsman W.~B., Sweigart A.~V., \& Grundahl, F. 2003, A\&A,
405, 135
\bibitem[2004]{mosw04}
Moehler, S., Sweigart, A.~V., Landsman, W.~B., Hammer, N.~J., \& Dreizler,
S. 2004, A\&A 415, 313 
\bibitem[1997]{napi97}
Napiwotzki, R. 1997, A\&A, 322, 256
\bibitem[2007]{nete07}
Newsham, G., Terndrup, D.~M. 2007, ApJ, 664, 332
\bibitem[2004]{norr04}
Norris, J. E. 2004, ApJ, 612, L25
\bibitem[2003]{orfe03}
Origlia, L., Ferraro, F.~R., Bellazzini, M., \& Pancino, E. 2003, ApJ, 591, 916
\bibitem[2005]{pivi05}
Piotto, G., Villanova, S., Bedin, L.~G., et al. 2005, ApJ, 621, 777 
\bibitem[2007]{pibe07} 
Piotto, G., Bedin L.~R., Anderson, J., et al. 2007, ApJ, 661, L53   
\bibitem[1997]{swei97} 
Sweigart, A.~V. 1997, The Third Conference on Faint
    Blue Stars, ed. A.\ G.\ D. Philip, J.\ Liebert \& R.\ A.\ Saffer
    (Schenectady: L. Davis Press), 3
\bibitem[2005]{ungl05}
Unglaub, K. 2005, The 14th European Workshop on White
Dwarfs, ASP Conf.\ Ser.\ Vol.\ 334,  eds.\ D.\ Koester \& S.\ Moehler (ASP:
San Francisco), p.\ 297
\bibitem[1999]{wedr99}
Werner, K., \& Dreizler, S. 1999,
The Journal of Computational and Applied Mathematics, Vol.\ 109,
eds. H. Riffert \& K. Werner, Elsevier Press, Amsterdam, p.\ 65
\bibitem[1998]{whro98} 
Whitney, J.~H., Rood, R.~T., O'Connell, R.~W., et al. 1998, ApJ, 495, 284 
\bibitem[2004]{zacharias2004}
Zacharias, N., Urban, S. E., Zacharias, M. I., Wycoff, G. L., 
Hall, D. M., Monet, D. G., Rafferty, T. J. 2004, AJ, 127, 3043
\end{thebibliography}
\end{document}